\documentclass[conference]{IEEEtran}
\IEEEoverridecommandlockouts
\usepackage{cite}
\usepackage{amsmath,amssymb,amsfonts}
\usepackage{algorithm}
\usepackage{algorithmicx}
\usepackage{graphicx}
\usepackage{textcomp}
\usepackage{xcolor}
\usepackage{fancyhdr}
\usepackage[hyphens]{url}

\usepackage{rotating}

\usepackage{graphicx}
\usepackage{comment}
\usepackage{multicol}
\usepackage{multirow}
\usepackage{booktabs} 
\usepackage{soul}
\usepackage{color, colortbl}
\usepackage{arydshln}

\usepackage[bookmarks=true,breaklinks=true,letterpaper=true,colorlinks,linkcolor=black,citecolor=blue,urlcolor=black]{hyperref}

\definecolor{Gray}{gray}{0.9}
\definecolor{LightCyan}{rgb}{0.88,1,1}

\newcolumntype{a}{>{\columncolor{Gray}}l}
\newcolumntype{d}{>{\columncolor{LightCyan}}l}

\usepackage{tikz}
\newcommand*\circled[1]{\tikz[baseline=(char.base)]{\node[fill=gray,shape=circle,draw,inner sep=0.6pt] (char) {#1};}}

\newcommand*\secondcircled[1]{\tikz[baseline=(char.base)]{\node[fill=white,shape=circle,draw,inner sep=0.6pt] (char) {#1};}}

\def\BibTeX{{\rm B\kern-.05em{\sc i\kern-.025em b}\kern-.08em
    T\kern-.1667em\lower.7ex\hbox{E}\kern-.125emX}}
\begin{document}

\title{HASI: Hardware-Accelerated Stochastic Inference, \\ A Defense Against Adversarial Machine Learning Attacks\thanks{This work was funded in part by the National Science Foundation under awards CCF-2028944 and CCF-1629392.}}

\author{
    \IEEEauthorblockN{Mohammad Hossein Samavatian, Saikat Majumdar, Kristin Barber, Radu Teodorescu}
    \IEEEauthorblockA{Department of Computer Science and Engineering \\ The Ohio State University, Columbus, OH, USA
    \\ \{samavatian.1, majumdar.42, barber.245, teodorescu.1\}@osu.edu}
}

\maketitle

\begin{abstract}
DNNs are known to be vulnerable to so-called adversarial attacks, in which inputs are carefully manipulated to induce misclassification. Existing defenses are mostly software-based and come with high overheads or other limitations. 
This paper presents HASI, a hardware-accelerated defense that uses a process we call stochastic inference to detect adversarial inputs. HASI carefully injects noise into the model at inference time and used the model's response to differentiate adversarial inputs from benign ones. 
We show an adversarial detection rate of average 87\% which exceeds the detection rate of the state of the art approaches, with a much lower overhead. We demonstrate  a software/hardware-accelerated co-design, which reduces the performance impact of stochastic inference to $1.58\times-2\times$ relative to the unprotected baseline, compared to $14\times-20\times$ overhead for a software-only GPU implementation. 
\end{abstract}

\section{Introduction}
\label{sec:introduction}

Deep neural networks (DNNs) are rapidly becoming indispensable tools for solving an increasingly diverse set of complex problems, including computer vision \cite{krizhevsky2012imagenet}, natural language processing \cite{collobert2008unified}, machine translation \cite{DBLP:journals/corr/BahdanauCB14}, and many others in different applications. 
Some of these application domains, such as medical, self-driving cars, face recognition, etc. require high accuracy outputs in order to gain public trust and widespread commercial adoption. Unfortunately, DNNs are known to be vulnerable to so-called "adversarial attacks" that purposefully compel classification algorithms to produce erroneous results. For example, in the computer vision domain, a large number of attacks 
\cite{carlini2017towards, moosavi-dezfooli2017universal, moosavi-dezfooli2016deepfool, goodfellow2015explaining, chen2018ead, kurakin2017adversarial, papernot2016the, madry2018towards} have demonstrated the ability to force state-of-the art classifiers 
to misclassify inputs that are carefully manipulated by an attacker. In most of the attacks, input images are only slightly altered such that they appear to the casual observer to be unchanged. However the alterations are made with sophisticated attacks that, in spite of the imperceptible changes to the input, result in reliable misclassification.  

\begin{figure}[htbp]
 \includegraphics[width=\linewidth]{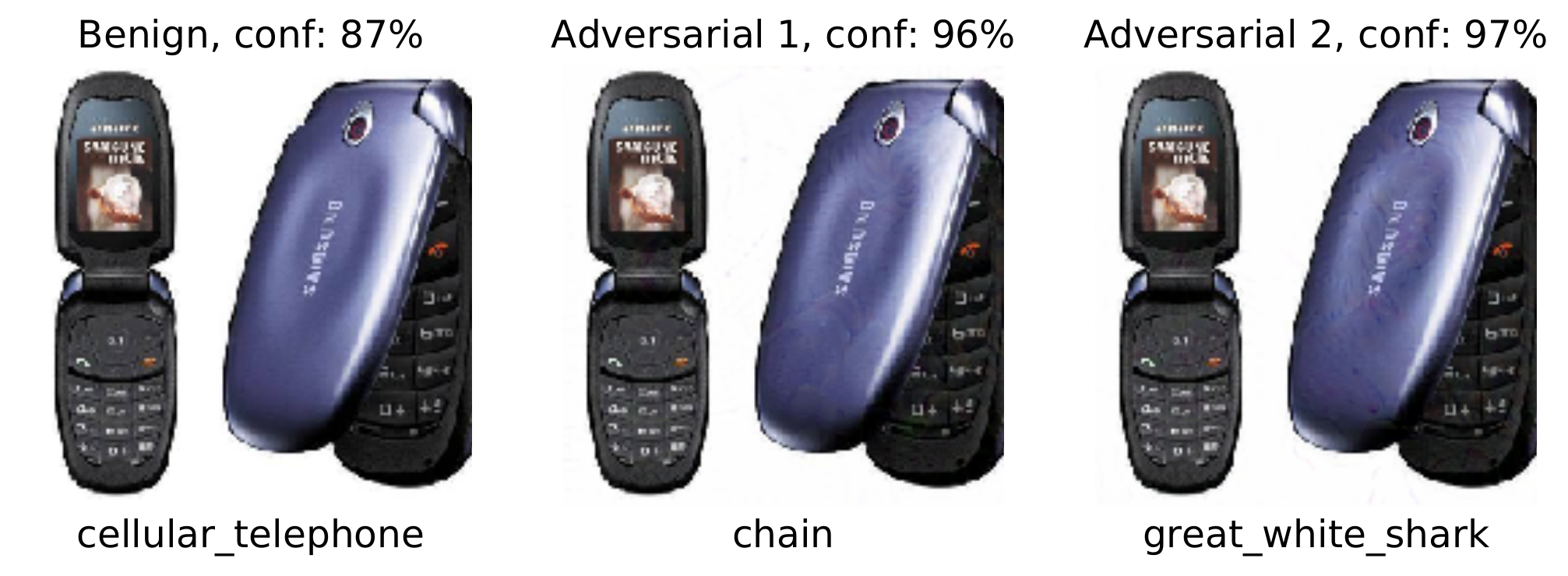}
\caption{\small Benign vs Adversarial Images.}\label{fig:benign_vs_adversarial}
\vspace{-1em}
\end{figure}

Figure \ref{fig:benign_vs_adversarial} shows an examples of adversarial images generated using the state-of-the art CW-$L_2$ attack \cite{carlini2017towards}. The leftmost top and bottom image is benign, unmodified samples that is correctly classified by a DNN (VGG16)
with 87\% confidence. The middle and rightmost pair of image represents the output of two versions of the CW-$L_2$ attack, each resulting in misclassification. Note that all adversarial images are virtually indistinguishable from the original to the casual observer, even though the confidence of the classifier is very high. 

Several defenses have been proposed to address adversarial attacks 
\cite{ma2019nic,papernot2016distillation,xu2018feature,dhillon2018stochastic,cao2017mitigating,ma2018characterizing}.
Many of these defenses follow approaches with the following limitations: (1) model hardening, which requires re-training and is not easily adaptable; (2) rely on certain characteristics of the dataset and do not generalize without profiling; and/or (3) have purely software implementations with potentially very high overheads, limiting their utility to real-world applications. Only a few hardware-accelerated defenses have been proposed. DNNGuard \cite{wang2020dnnguard} relies on a separate dedicated classifier for detection, which also comes at a high cost. \cite{guesmi2020defensive} takes advantage of approximation to make the models more robust. However the  approach is not adaptive and the source of the noise within the approximation is input dependent, which makes the defense easily vulnerable to adaptive attacks.

This paper presents HASI, a hardware/software co-designed defense that relies on a novel stochastic inference process to effectively defend against state-of-the art adversarial attacks. 
By carefully injecting noise into the model during inference, we can differentiate adversarial inputs from benign ones. 
HASI is a dynamic approach in which noise is injected throughout the model during inference. HASI is the first work to recognize the correlation between classification confidence and the probability that an adversarial image will be detected. Building on this observation, HASI uses adaptive noise injection, increasing the level of injected noise for inputs with high confidence classifications. This results in robust detection across a broad set of attacks with high accuracy and low false positive rate. 
We show an adversarial detection rate of 86\% when applied to VGG16 and 88\% when applied to ResNet50, which exceeds the detection rate of the state of the art approaches. We also show that HASI is robust against attacks that are aware of the defense and attempt to circumvent it. 

In order to reduce the performance impact, we propose an
accelerator design we call a \textbf{Dy}namically \textbf{S}parsified \textbf{CNN} (\textbf{DySCNN}). In this design, noise is introduced into the network by randomly dropping weights from the model, effectively sparsifying the network. The degree of sparsification is dynamically changed based on the confidence of the initial classification. 
The HASI hardware/software co-design reduces the performance impact of stochastic inference to $1.58\times-2\times$ relative to the unprotected baseline, compared to $14\times-20\times$ overhead for a software-only GPU implementation.

\section{Threat Model}
\label{sec:threat}

We assume in this paper that the adversary has complete access to the network, including the output prediction and logits, with full knowledge of the architecture and parameters, and is able to use this in a white-box manner. 
We focus mainly on recent state-of-the-art optimization-based attacks \textbf{CW} \cite{carlini2017towards} and \textbf{EAD} \cite{chen2018ead} since it has been demonstrated that all earlier attacks can be overcome utilizing other methods, such as adversarial training \cite{goodfellow2014explaining} 
or defensive distillation \cite{papernot2016distillation}, 
which could be used in combination with our approach.
Additionally, we verify our evaluation includes high confidence adversarial examples, as some previously proposed defenses were later shown to perform poorly under a more holistic treatment which included these inputs \cite{lu2018on}.

\section{Background}
\label{sec:background}

\subsection{Adversarial Attacks}

Adversarial attacks were first introduced by Szegedy et al. in \cite{szegedy2014intriguing}, which showed that despite the high accuracy of machine learning models, small perturbations to inputs can reliably force misclassifications--while the perturbed input remains indistinguishable from the original seed to the naked eye. 
The objective of an adversarial attack is to force the output classification for some maliciously crafted input \textbf{x$^\prime$}, based on a benign input \textbf{x}, to be incorrect with respect to \textbf{x}.  Attacks can be \emph{targeted}, where the adversary's goal is for \textbf{x$^\prime$} to be misclassified as a particular class \emph{t}, or \emph{untargeted}, such that a misclassification of \textbf{x$^\prime$} to any class other than the correct class of \textbf{x} (ground truth) is sufficient. 

\subsection{Robustness in Neural Networks}

Robustness can be informally defined as the measure of how difficult it is to find adversarial examples close to their original inputs.
Several methods for designing robust neural networks to adversarial attacks have been proposed in the literature.  These methods typically fall into four broad categories \cite{akhtar2018threat}: 1) \textbf{hardening the model}, 2) \textbf{hardening the test inputs}, 3) \textbf{adding a secondary-external network}, and 4) \textbf{modifying the network post-training}.  

Unlike model and input hardening approaches \cite{papernot2016distillation}, HASI does \emph{not require any re-training or input pre-processing} of the model and sacrifices little in model accuracy. HASI is designed to \emph{detect} adversarial examples post-training, during model inference. HASI does require multiple inference passes, but data reuse optimizations help to alleviate this overhead.
HASI noise injection allows for detection to be more easily generalized, as opposed to other methods which require profiling to select appropriate parameters, such as Feature Squeezing \cite{xu2018feature} and Path Extraction \cite{qiu2019adversarial,9251936}. 
\section{Impact of Model Noise on Classification Output}
\label{sec:noise_impact}

Prior work such as \cite{hu2019a,cohen2019certified,lecuyer2019certified,roth2019odds} have shown that adding some amount of random noise to images can help DNNs correctly classify adversarial inputs. The hypothesis advanced by prior work is that high-quality adversarial inputs that have low distortion occur with low probability, which means they reside in small and low-density pockets of the classification space \cite{ma2018characterizing,szegedy2013intriguing}. Therefore injecting noise into inputs has a high probability of moving adversarial inputs out of the low density pockets into the correct classification. At the same time, an equal amount of noise is less likely to result in misclassification of benign inputs. 

In this work we build on these prior observations to construct the proposed defense. However, instead of focusing on correctly classifying adversarial inputs, we instead build a mechanism for detecting them with higher accuracy. The key novel observations we make in this work is that injecting noise in the model - rather than the input - and adapting the amount of noise to the confidence of the classification improves the ability to detect stronger adversarial inputs, for which correct classification cannot be achieved. 

\begin{figure}[htb]
\centering
 \includegraphics[width=0.9\linewidth]{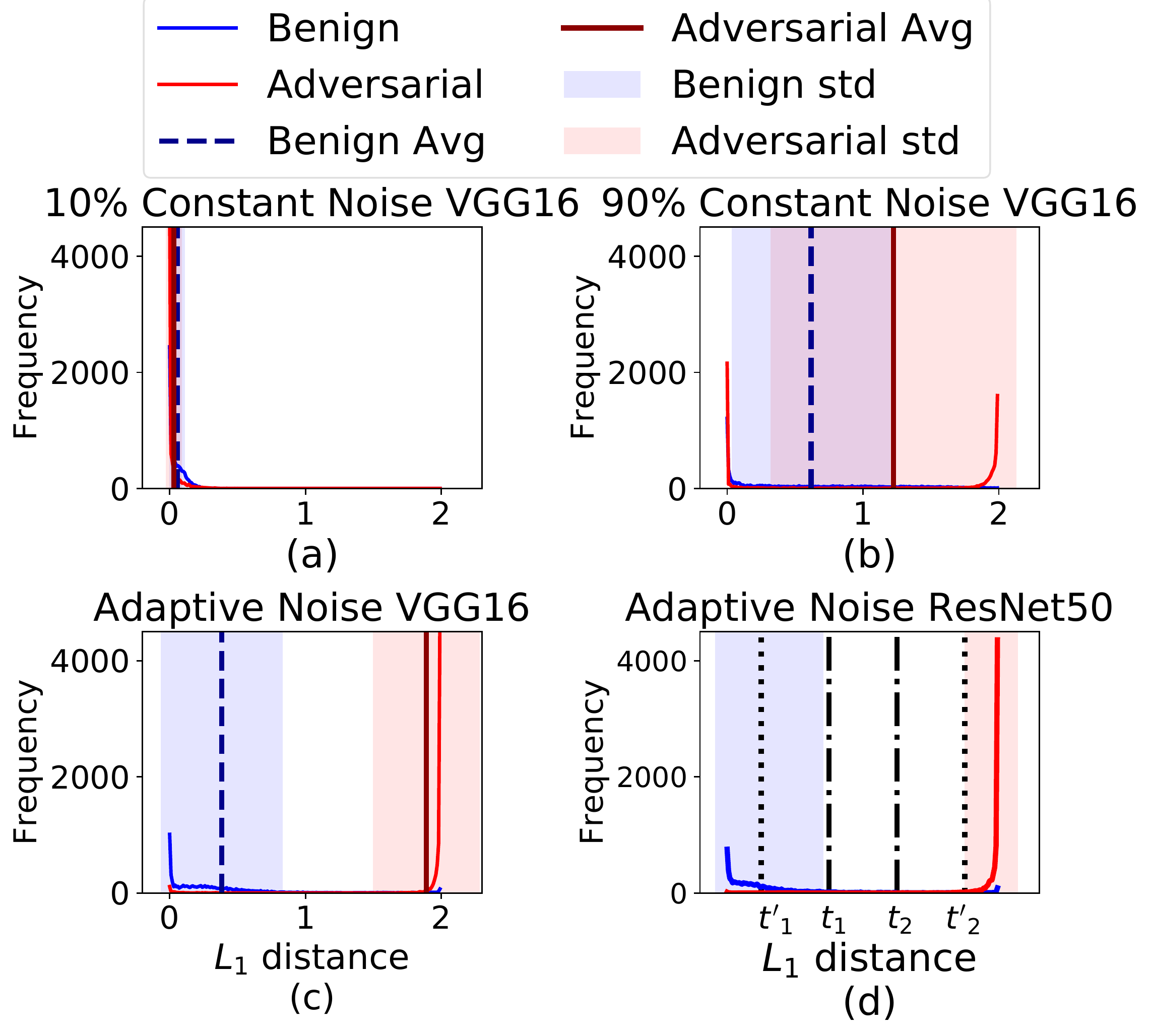}
\caption{\small $L_1$ distance distributions for benign and adversarial images under varying levels of noise, for VGG16 (and ResNet50).}\label{Fig.:distance}
\vspace{-1em}
\end{figure}

To achieve this goal, we focus on the impact of model noise injection on the probability distributions of the classification outputs. We use the $L_1$-norm metric to measure the impact of model noise on inference output probability distributions. The $L_1$-norm 
is the sum of the absolute pair-wise differences between elements of two vectors. 
We measure the $L_1$-norm difference between the output probability distribution vectors of noisy and non-noisy runs of the same inputs. Figure \ref{Fig.:distance} shows the $L_1$ distance distributions for sample of 1000 benign and adversarial images. 

Figure \ref{Fig.:distance} (a) and (b) shows the $L_1$ distance distributions for benign (blue line) and adversarial (red line) inputs, when 10\% and 90\% noise is injected in all activation layers. 
Each data point represents the $L_1$ distance between the output vector of a noisy run and that of a non-noisy run of the {\bf same input}. Mean and one standard deviation from the mean is also highlighted. 
We can see that the two distributions (benign and adversarial) are almost identical, with low $L_1$ distance values for 10\% noise. For 90\% noise the $L_1$ distance values shift to the right and still it cannot separate two distribution with clear margins. 


Finally, Figures \ref{Fig.:distance} (c) and (d) show the $L_1$ distance distributions for an adaptive noise injection approach in which we vary the amount of noise injected as a function of the confidence of the classification.  The higher the classification confidence, the higher the level of noise we inject into the model. Tailoring the level of noise to the confidence helps detect well-trained high confidence adversarials, while reducing the likelihood of misclassification of benign images. The adaptive noise is the most successful at generating a large distance between distributions of benign and adversarial images, improving detection accuracy. As a result, we use adaptive noise injection in the HASI framework.

\section{HASI Stochastic Inference}
\label{sec:design}

The HASI framework relies on the $L_1$ distance observation to detect adversarial inputs. Initially, a first inference pass through the network is performed without noise injection to establish a reference. The output classification is recorded as $P^b$. This is followed by another inference pass, this time with noise injected into the model, with output $P^{NSpr}$. The confidence of the classification $P^b$ is used to determine the amount of noise to be injected. Next, the $L_1$ distance between the output vectors of the noisy ($P^{NSpr}$) and non-noisy ($P^b$) inference passes is computed and compared with detection threshold values. If the measured $L_1$ distance is either very high or very low ($<t'_1$ or $>t'_2$ in  Figure \ref{Fig.:distance}-d) the input image can immediately
be classified as benign or adversarial, respectively.
Otherwise, HASI cannot yet make a high-confidence detection, and another noisy inference pass is required. The average $L_1$ distance over all the previous noisy runs is computed and compared with the more conservative thresholds $t_1$ and $t_2$ for detection. 
\subsection{Noisy Sparsification}
\label{noisy_sparsification}

In order to reduce the performance impact of the second inference pass, we use dynamic model sparsification to introduce model noise. This process, called "noisy sparsification", introduces noise into the DNN by randomly "dropping" (essentially ignoring) weights from the model. The fraction of dropped weights ("sparsification rate") controls the amount of noise.
The sparsification rate for each input is determined based on the classification confidence of the non-noisy run of that input.

The main advantage of Noisy Sparsification 
is the potential reduction in performance overhead of noisy inference. However, exploiting sparse filters to reduce computation time is challenging because of the workload imbalance across otherwise homogeneous compute units. This is even more challenging in the case of  $HASI^{NSpr}$ because the filter sparsity changes randomly from run to run. 

Since the sparsification is dynamically changing most of the designs are not efficient to leverage the sparse computation in HASI. 
In order to address the challenges of dynamic sparsification in HASI we developed a hardware-software co-designed accelerator for a dynamically sparsified  model, based on the TensorDash \cite{mahmoud2020tensordash} architecture. We call our design a Dynamic sparsified CNN (DySCNN) accelerator.

DySCNN consists of two components: 1) software scheduler and 2) hardware accelerator. 

\begin{figure}[htbp]
  \centering
     \includegraphics[width=0.9\linewidth]{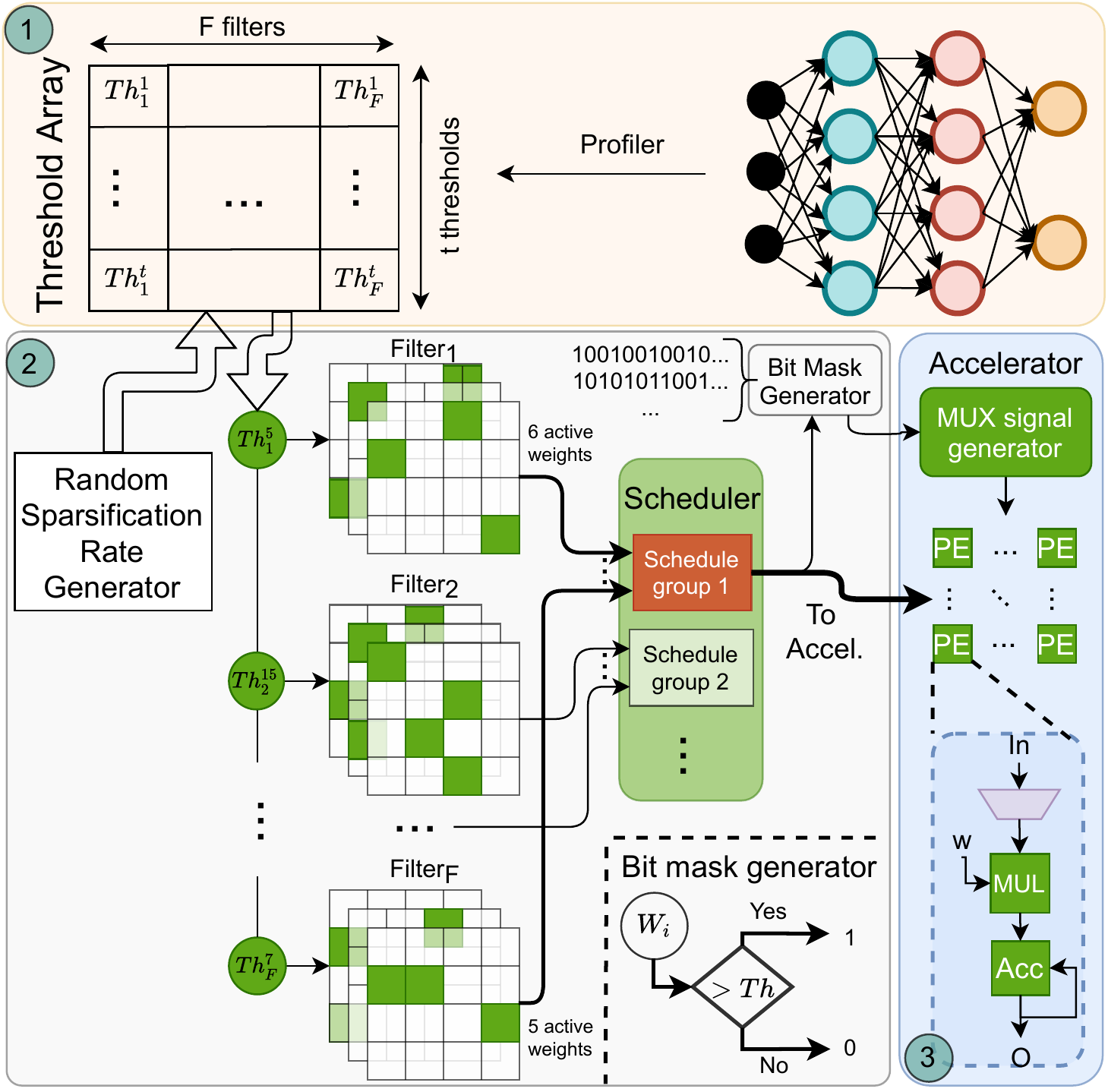}
  \caption{\small DySCNN software scheduler and hardware acceleration.}\label{Fig.:DySCNN_driver}
\end{figure}

The software scheduler (Figure \ref{Fig.:DySCNN_driver}) handles two main tasks: I) one-time offline profiling (\circled{1}) and II) online scheduling(\circled{2}). In order to reduce the overhead of online sparsification, the DySCNN Scheduler parses the model offline and generates a table of threshold values for each filter, corresponding to different sparsification rates. The threshold values will be used by the scheduler to determine which weights to be dropped based on the sparsification rate that randomly assigned to each filter at run time. 

At run time the sparsification rate generator issues sparsification rates for each kernel and retrieves the corresponding threshold values. Once a threshold is assigned all weigths with values below that threshold will be ignored. The scheduler then selects filters with similar numbers of active weights and groups them before deployment on the accelerator.
This will increase the efficiency of the inference run 
by reducing load imbalance and the number of idle cycles. The Scheduler also creates bit masks which represent active and dropped weights. The bit mask will be used by a MUX signal generator (\circled{3}) in the accelerator hardware to map the correct inputs to the active weights. We used a look-ahead mechanism similar to that in \cite{mahmoud2020tensordash} to match inputs and weights. However, DySCNN reduces the complexity of the design in \cite{mahmoud2020tensordash} because it relies on the software scheduler to balance the number of active weights across compute lanes. Note that only active weights will be loaded into the weight buffers of the accelerator memory; dropped weights will be ignored, saving memory energy.

\section{Evaluation} 
\label{sec:eval}

\subsection{Methodology} 
\label{sec:method}
As a proof-of-concept, we implement HASI in a FPGA-based DNN accelerator, the Xilinx CHaiDNN architecture \cite{chaidnn}. We modified CHaiDNN software stack to include the HASI API. 
We synthesize and deploy CHaiDNN+HASI on a Xilinx Zynq UltraScale+ FPGA. 
We compare our FPGA accelerator with software implementation of HASI on a CPU and GPU using TensorFlow2.2 \cite{tensorflow}.
We run our software HASI on Intel Core-i7 CPU@3.40GHz and NVIDIA RTX-2060 Turing GPU . 
We used two networks VGG16 \cite{simonyan2014deep} and ResNet50 \cite{ResNet} trained on ImageNet \cite{krizhevsky2012imagenet} for running attacks and generating adversarial images. 
Attacks aim to misclassify an input into a target class, use two types of targets 1) \textit{Next} likely class and 2) least likely class (\textit{LL}). 
Table \ref{tab:attack_stats} summarizes the adversarial attacks alongside their detailed parameters and average $L_2$ distortion. 
We compared our detection rate of adversarial as well as True Positive rate (TPR) of benign images with \textbf{Feature squeezing (FS)}\cite{xu2018feature}
which is a correction-detection mechanism that relies on reducing the input space (and attack surface) by "squeezing" images.
FS requires off-line profiling and training to find the best squeezer and corresponding thresholds for each pair of data-set and attack, making it less practical to deploy in real-world applications.

\begin{table}[htbp]
\vspace{-1em}
\caption{Attack parameters of CW and EAD attacks. Attacks detection rates for defenses FS and HASI. 
}
\label{tab:attack_stats}
\small
\centering
\scalebox{0.75}{
\begin{tabular}{|l|c|ad|ad||ad|ad|}
\hline
\multicolumn{6}{|c||}{\textbf{Attacks}} & \multicolumn{4}{c|}{\textbf{Defenses (detection \%)}}\\
\hline\hline  
\multirow{2}{*}{ Attack}&\multirow{2}{*}{ Target} &\multicolumn{2}{c|}{Param k } & \multicolumn{2}{c||}{ $L_2$ Distortion}  & \multicolumn{2}{c|}{ FS$^+$} & \multicolumn{2}{c|}{ $HASI^{NSpr}$} \\
\cline{3-10}
       && VG & RN  &  VGG & RNet & VGG & RNet & VGG & RNet\\\hline

\multirow{2}{*}{\textbf{$CW_{L0}$} } &    Next     & 5& 5  & 10.44 &           6.26 &        100  &     100   &67  & 82 \\

            & LL &      & &                13.65 &           8.33  &  100  &     100  &   100  &100 \\
          
\hline

\multirow{4}{*}{\textbf{$CW_{L2}$}}&  Next & 10 & 30 &       1.69 &           1.63 &               84  &      89  & 91  & 100 \\

            &  LL &   &      &         2.24 &           2.07 &   100  &     100  &  100  & 100 \\
            
\cdashline{2-10}
          & Next &70&140&    7.43 &           9.22 &  6  &      48   &84  & 89 \\

            &LL  &    &      &                  8.07 &          11.56 &      9  &      67  &  96  &98 \\

\hline
\multirow{2}{*}{\textbf{$CW_{L\infty}$}}   &  Next & 5 & 5&           2.27 &            1.6  &   91  &      96  &  83  & 89 \\

     & LL &    &       &                3.05 &           2.12 &   100  &     100   &100  & 100 \\
\hline

\multirow{4}{*}{\textbf{$EAD_{L1}$}} &Next &10&30&             2.69 &           2.79 &  78  &      98  & 91  & 98 \\

   & LL  &   &  &            3.56 &           3.47  &  100  &      98  &  100  &     100 \\
   
\cdashline{2-10}
 & Next & 70& 140&                  9.92 &          12.34 & 4  &      45  & 78   & 88 \\

  & LL  &   &    &               10.65 &          17.36 &     4  &      81   & 93  & 84 \\

\hline
\multirow{2}{*}{\textbf{$EAD_{EN}$}}& Next & 10& 30 &             4.36 &           6.73 &         63  &      89  & 80  & 94 \\

    & LL  &   &     &          5.9 &           8.59 &    98  &      98  & 99  & 97 \\

\hline
\multicolumn{5}{|l}{RNet: ResNet50, VGG: VGG16}&
\multicolumn{1}{r||}{AVG}&     55  &       79   &  86  & 88 \\
\cline{6-10}

\multicolumn{5}{|l}{$^+$threshold: VGG: 1.022, ResNet: 1.229}&\multicolumn{1}{r||}{FPR}&    6  &       3  &  6  & 6 \\

\hline
\end{tabular}
}
\vspace{-1em}
\end{table}


\subsection{Adversarial Detection Rate} 
Table \ref{tab:attack_stats} lists the detection rate for all the attack variants we evaluate, for both VGG and ResNet. 
The results show that 
$HASI^{NSpr}$ outperform 
FS 
on average. $HASI^{NSpr}$ shows an average detection rate of 
86\% and 88\% for VGG and ResNet, respectively. $HASI^{NSpr}$ significantly outperforms the state of the art defense, FS which averages 55\% and 79\% for VGG and ResNet, respectively. $HASI^{NSpr}$ is especially strong at detecting more recent, high-confidence attacks, for which FS does not work as well.
Prior work has shown that increasing the factor {\em k} improves the strength of the adversarials at the cost of somewhat higher input distortion. For instance, under the $EAD_{L1}$ attack with $k=70$ we see 93\% detection rates with the HASI designs vs. 4\% for FS (VGG16). This shows that HASI is resilient to very strong attacks which still have acceptable quality with minor visible changes to the image. See Appendix \ref{strength_analysis} for additional analysis on attack strength. 

\subsection{HASI-Aware Attacks}

We also examine a modified attack that assumes knowledge of the HASI design and is optimized to defeat it. 
We used 
the approach suggested in \cite{tramer2020adaptive} to generate adversarial examples (details in Appendix \ref{adaptive_attack_append}).
Table \ref{tab:adaptive} summarizes the adaptive attack parameters and detection rate under HASI. $\beta$ is regularization that tune the trade-off between $L_2$ distortion and $L_1$ distance. We can see that for low-$\beta$ attacks, HASI detection rate is very high ($A_{1}$-$A_{3}$). For $\beta=10^{-1}$ the detection rate is lower, but still acceptable for VGG. This shows HASI is resilient to HASI-aware attacks that optimize for low $L_1$ distance.  

\begin{table}[htbp]
\vspace{-1em}
\centering
\caption{HASI aware adaptive attacks} 
\label{tab:adaptive}
\scalebox{0.9}{
\begin{tabular}{|c|c|ad|ad|ad|}
\hline
\multirow{2}{*}{Attack} & \multirow{2}{*}{$\beta$}  & \multicolumn{2}{c}{Mean Confidence} & \multicolumn{2}{|c}{$L_2$ Distortion}   & \multicolumn{2}{|c|}{HASI detection}  \\
\cline{3-8}
 &             &     VGG                 &    RNet                   &  VGG   &       RNet      &       VGG        &    RNet             \\
\hline
 $A_1$  & $10^{-4}$     &                 94.3\% &                    95.9\% &        3.91 &           2.71 &   99\% &     100\% \\
   
$A_2$ &$10^{-3}$      &                 92.3\% &                    94.0\% &        2.48 &           1.42 &   99\% &     100\% \\
 
$A_3$ &$10^{-2}$       &                 96.1\% &                    96.9\% &       10.14 &           8.45 &   95\% &      84\% \\
 
 $A_4$  &    $10^{-1}$  &                 97.7\% &                    97.9\% &       41.58 &          46.99 &   81\% &      39\% \\
\hline
\end{tabular}
}
\vspace{-1em}
\end{table}


\subsection{Performance Overhead}
\begin{figure}[htbp]
  \centering
    \includegraphics[width=\linewidth]{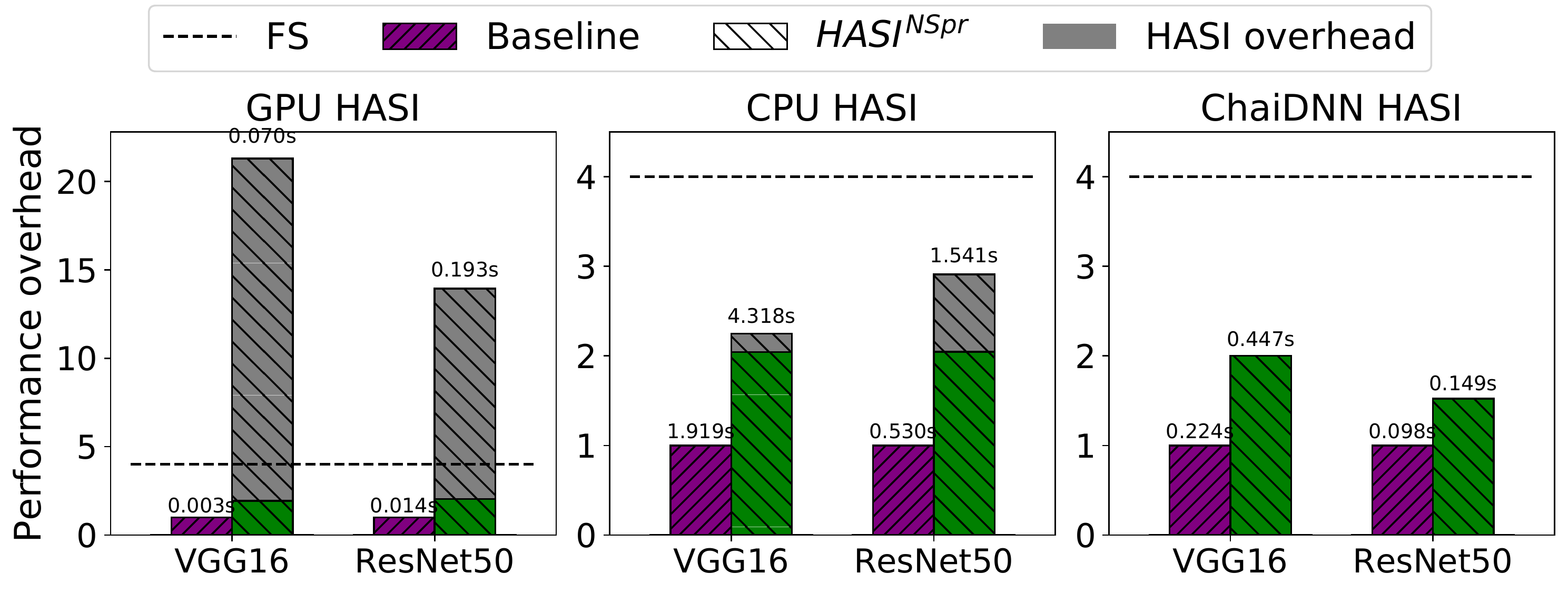}
  \caption{\small Normalized run time of HASI on GPU, CPU and CHaiDNN FGPA-based accelerator design for VGG16 and ResNet50 v.s. FS.}\label{Fig.:runtime}
  \vspace{-1em}
\end{figure}

We next examine the performance 
overhead of the HASI design. Figure \ref{Fig.:runtime} shows the average normalized run time of  HASI on different platforms. The runtime overhead of HASI on GPU is $6\times$ to $20\times$ higher than the baseline, due to the inefficiency of dynamic sparsification. For the CPU these numbers are $2.25\times$ and $3\times$. The noisy sparsification accelerator reduces the overhead to $1.99\times$ and $1.6\times$ for VGG16 and ResNet50, respectively.
FS also requires at least 3 squeezers, resulting in at least 4$\times$ performance overhead. 

\section{Conclusion}
\label{sec:conclusion}

In conclusion, this paper showed that adaptive noise injection in DNN models enables robust $>90\%$ adversarial detection across multiple strong attacks, for different image classifiers. We also showed HASI's robustness against attacks that are aware of the defense and attempt to circumvent it. We demonstrated a hardware/software co-designs of HASI to reduce the performance impact of stochastic inference to $1.6\times$ and $1.99\times$ for ResNet50 and VGG16 respectively.

\bibliographystyle{IEEEtranS}
\bibliography{main}

\begin{thebibliography}{10}
\providecommand{\url}[1]{#1}
\csname url@samestyle\endcsname
\providecommand{\newblock}{\relax}
\providecommand{\bibinfo}[2]{#2}
\providecommand{\BIBentrySTDinterwordspacing}{\spaceskip=0pt\relax}
\providecommand{\BIBentryALTinterwordstretchfactor}{4}
\providecommand{\BIBentryALTinterwordspacing}{\spaceskip=\fontdimen2\font plus
\BIBentryALTinterwordstretchfactor\fontdimen3\font minus
  \fontdimen4\font\relax}
\providecommand{\BIBforeignlanguage}[2]{{%
\expandafter\ifx\csname l@#1\endcsname\relax
\typeout{** WARNING: IEEEtranS.bst: No hyphenation pattern has been}%
\typeout{** loaded for the language `#1'. Using the pattern for}%
\typeout{** the default language instead.}%
\else
\language=\csname l@#1\endcsname
\fi
#2}}
\providecommand{\BIBdecl}{\relax}
\BIBdecl

\bibitem{tensorflow}
\BIBentryALTinterwordspacing
``Tensorflow,'' \url{https://www.tensorflow.org/}. [Online]. Available:
  \url{https://www.tensorflow.org/}
\BIBentrySTDinterwordspacing

\bibitem{akhtar2018threat}
N.~{Akhtar} and A.~{Mian}, ``Threat of adversarial attacks on deep learning in
  computer vision: A survey,'' \emph{IEEE Access}, vol.~6, pp.
  14\,410--14\,430, 2018.

\bibitem{DBLP:journals/corr/BahdanauCB14}
D.~Bahdanau, K.~Cho, and Y.~Bengio, ``Neural machine translation by jointly
  learning to align and translate,'' in \emph{3rd International Conference on
  Learning Representations ({ICLR})}, 2015.

\bibitem{cao2017mitigating}
X.~{Cao} and N.~Z. {Gong}, ``Mitigating evasion attacks to deep neural networks
  via region-based classification,'' in \emph{Proceedings of the 33rd Annual
  Computer Security Applications Conference on}, 2017, pp. 278--287.

\bibitem{carlini2017towards}
N.~{Carlini} and D.~{Wagner}, ``Towards evaluating the robustness of neural
  networks,'' in \emph{IEEE Symposium on Security and Privacy (SP)}, 2017.

\bibitem{chen2018ead}
P.-Y. Chen, Y.~Sharma, H.~Zhang, J.~Yi, and C.-J. Hsieh, ``Ead: elastic-net
  attacks to deep neural networks via adversarial examples,'' in
  \emph{Thirty-second AAAI conference on artificial intelligence}, 2018.

\bibitem{cohen2019certified}
J.~M. Cohen, E.~Rosenfeld, and J.~Z. Kolter, ``Certified adversarial robustness
  via randomized smoothing,'' \emph{arXiv preprint arXiv:1902.02918}, 2019.

\bibitem{collobert2008unified}
R.~Collobert and J.~Weston, ``A unified architecture for natural language
  processing: Deep neural networks with multitask learning,'' in
  \emph{Proceedings of the 25th international conference on Machine learning
  ({ICLR})}, 2008, pp. 160--167.

\bibitem{dhillon2018stochastic}
G.~S. {Dhillon}, K.~{Azizzadenesheli}, J.~D. {Bernstein}, J.~{Kossaifi},
  A.~{Khanna}, Z.~C. {Lipton}, and A.~{Anandkumar}, ``Stochastic activation
  pruning for robust adversarial defense,'' in \emph{International Conference
  on Learning Representations (ICLR)}, 2018.

\bibitem{9251936}
Y.~{Gan}, Y.~{Qiu}, J.~{Leng}, M.~{Guo}, and Y.~{Zhu}, ``Ptolemy: Architecture
  support for robust deep learning,'' in \emph{2020 53rd Annual IEEE/ACM
  International Symposium on Microarchitecture (MICRO)}, 2020, pp. 241--255.

\bibitem{goodfellow2014explaining}
I.~J. Goodfellow, J.~Shlens, and C.~Szegedy, ``Explaining and harnessing
  adversarial examples,'' \emph{arXiv preprint arXiv:1412.6572}, 2014.

\bibitem{goodfellow2015explaining}
I.~J. {Goodfellow}, J.~{Shlens}, and C.~{Szegedy}, ``Explaining and harnessing
  adversarial examples,'' in \emph{ICLR 2015 : International Conference on
  Learning Representations 2015}, 2015.

\bibitem{guesmi2020defensive}
A.~Guesmi, I.~Alouani, K.~Khasawneh, M.~Baklouti, T.~Frikha, M.~Abid, and
  N.~Abu-Ghazaleh, ``Defensive approximation: Enhancing cnns security through
  approximate computing,'' \emph{arXiv preprint arXiv:2006.07700}, 2020.

\bibitem{ResNet}
K.~{He}, X.~{Zhang}, S.~{Ren}, and J.~{Sun}, ``Deep residual learning for image
  recognition,'' in \emph{2016 IEEE Conference on Computer Vision and Pattern
  Recognition (CVPR)}, 2016, pp. 770--778.

\bibitem{hu2019a}
S.~{Hu}, T.~{Yu}, C.~{Guo}, W.-L. {Chao}, and K.~{Weinberger}, ``A new defense
  against adversarial images: Turning a weakness into a strength,'' in
  \emph{NeurIPS 2019 : Thirty-third Conference on Neural Information Processing
  Systems}, 2019, pp. 1633--1644.

\bibitem{krizhevsky2012imagenet}
A.~Krizhevsky, I.~Sutskever, and G.~E. Hinton, ``Imagenet classification with
  deep convolutional neural networks,'' in \emph{Advances in neural information
  processing systems}, 2012, pp. 1097--1105.

\bibitem{kurakin2017adversarial}
A.~{Kurakin}, I.~J. {Goodfellow}, and S.~{Bengio}, ``Adversarial examples in
  the physical world,'' in \emph{International Conference on Learning
  Representations (ICLR)}, 2017.

\bibitem{lecuyer2019certified}
M.~Lecuyer, V.~Atlidakis, R.~Geambasu, D.~Hsu, and S.~Jana, ``Certified
  robustness to adversarial examples with differential privacy,'' in \emph{2019
  IEEE Symposium on Security and Privacy (SP)}.\hskip 1em plus 0.5em minus
  0.4em\relax IEEE, 2019, pp. 656--672.

\bibitem{lu2018on}
P.-H. {Lu}, P.-Y. {Chen}, and C.-M. {Yu}, ``On the limitation of local
  intrinsic dimensionality for characterizing the subspaces of adversarial
  examples,'' in \emph{ICLR (Workshop)}, 2018.

\bibitem{ma2019nic}
S.~{Ma}, Y.~{Liu}, G.~{Tao}, W.-C. {Lee}, and X.~{Zhang}, ``Nic: Detecting
  adversarial samples with neural network invariant checking.'' in \emph{NDSS},
  2019.

\bibitem{ma2018characterizing}
X.~{Ma}, B.~{Li}, Y.~{Wang}, S.~M. {Erfani}, S.~{Wijewickrema}, M.~E. {Houle},
  G.~{Schoenebeck}, D.~{Song}, and J.~{Bailey}, ``Characterizing adversarial
  subspaces using local intrinsic dimensionality,'' \emph{arXiv preprint
  arXiv:1801.02613}, 2018.

\bibitem{madry2018towards}
A.~{Madry}, A.~{Makelov}, L.~{Schmidt}, D.~{Tsipras}, and A.~{Vladu}, ``Towards
  deep learning models resistant to adversarial attacks,'' in
  \emph{International Conference on Learning Representations (ICLR)}, 2018.

\bibitem{mahmoud2020tensordash}
M.~Mahmoud, I.~Edo, A.~H. Zadeh, O.~M. Awad, G.~Pekhimenko, J.~Albericio, and
  A.~Moshovos, ``Tensordash: Exploiting sparsity to accelerate deep neural
  network training,'' in \emph{2020 53rd Annual IEEE/ACM International
  Symposium on Microarchitecture (MICRO)}.\hskip 1em plus 0.5em minus
  0.4em\relax IEEE, 2020, pp. 781--795.

\bibitem{moosavi-dezfooli2017universal}
S.-M. {Moosavi-Dezfooli}, A.~{Fawzi}, O.~{Fawzi}, and P.~{Frossard},
  ``Universal adversarial perturbations,'' in \emph{2017 IEEE Conference on
  Computer Vision and Pattern Recognition (CVPR)}, 2017, pp. 86--94.

\bibitem{moosavi-dezfooli2016deepfool}
S.-M. {Moosavi-Dezfooli}, A.~{Fawzi}, and P.~{Frossard}, ``Deepfool: A simple
  and accurate method to fool deep neural networks,'' in \emph{IEEE Conference
  on Computer Vision and Pattern Recognition (CVPR)}, 2016.

\bibitem{papernot2016distillation}
N.~{Papernot}, P.~{McDaniel}, X.~{Wu}, S.~{Jha}, and A.~{Swami}, ``Distillation
  as a defense to adversarial perturbations against deep neural networks,'' in
  \emph{2016 IEEE Symposium on Security and Privacy (SP)}, 2016, pp. 582--597.

\bibitem{papernot2016the}
N.~{Papernot}, P.~D. {McDaniel}, S.~{Jha}, M.~{Fredrikson}, Z.~B. {Celik}, and
  A.~{Swami}, ``The limitations of deep learning in adversarial settings,'' in
  \emph{IEEE European Symposium Security and Privacy}, 2016.

\bibitem{qiu2019adversarial}
Y.~Qiu, J.~Leng, C.~Guo, Q.~Chen, C.~Li, M.~Guo, and Y.~Zhu, ``Adversarial
  defense through network profiling based path extraction,'' in
  \emph{Proceedings of the IEEE Conference on Computer Vision and Pattern
  Recognition}, 2019, pp. 4777--4786.

\bibitem{roth2019odds}
K.~Roth, Y.~Kilcher, and T.~Hofmann, ``The odds are odd: {A} statistical test
  for detecting adversarial examples,'' in \emph{{ICML}}, ser. Proceedings of
  Machine Learning Research, vol.~97, 2019, pp. 5498--5507.

\bibitem{simonyan2014deep}
K.~Simonyan and A.~Zisserman, ``Very deep convolutional networks for
  large-scale image recognition,'' 2014.

\bibitem{szegedy2013intriguing}
C.~Szegedy, W.~Zaremba, I.~Sutskever, J.~Bruna, D.~Erhan, I.~Goodfellow, and
  R.~Fergus, ``Intriguing properties of neural networks,'' \emph{arXiv preprint
  arXiv:1312.6199}, 2013.

\bibitem{szegedy2014intriguing}
C.~{Szegedy}, W.~{Zaremba}, I.~{Sutskever}, J.~{Bruna}, D.~{Erhan},
  I.~{Goodfellow}, and R.~{Fergus}, ``Intriguing properties of neural
  networks,'' in \emph{International Conference on Learning Representations
  2014}, 2014.

\bibitem{tramer2020adaptive}
F.~Tramer, N.~Carlini, W.~Brendel, and A.~Madry, ``On adaptive attacks to
  adversarial example defenses,'' \emph{arXiv preprint arXiv:2002.08347}, 2020.

\bibitem{wang2020dnnguard}
\BIBentryALTinterwordspacing
X.~Wang, R.~Hou, B.~Zhao, F.~Yuan, J.~Zhang, D.~Meng, and X.~Qian, ``Dnnguard:
  An elastic heterogeneous dnn accelerator architecture against adversarial
  attacks,'' in \emph{Proceedings of the Twenty-Fifth International Conference
  on Architectural Support for Programming Languages and Operating Systems},
  ser. ASPLOS ’20.\hskip 1em plus 0.5em minus 0.4em\relax New York, NY, USA:
  Association for Computing Machinery, 2020, p. 19–34. [Online]. Available:
  \url{https://doi.org/10.1145/3373376.3378532}
\BIBentrySTDinterwordspacing

\bibitem{chaidnn}
Xilinx, ``{CHaiDNN},'' https://github.com/Xilinx/CHaiDNN.

\bibitem{xu2018feature}
W.~{Xu}, D.~{Evans}, and Y.~{Qi}, ``Feature squeezing: Detecting adversarial
  examples in deep neural networks.'' in \emph{Proceedings 2018 Network and
  Distributed System Security Symposium}, 2018.

\end{thebibliography}
\appendix
\section{Appendix}
\subsection{Adaptive Attack Design Details}
\label{adaptive_attack_append}
In HASI aware attack, for sample x of class y, we pick a target $t\neq y$ and create adversarial example $x'$ that minimize the objective:
\begin{equation} 
    \textrm{$minimize$}~~~||y(x')-y(x_t)||_1
\end{equation}
where $y(x')$ and $y(x_t)$ are the probability vector of adversarial and target input respectively. While we try to minimize the $L_1$ distance between adversarial and the benign target, we need to also minimize the adversarial perturbation under $L_2$  distortion metric. The final objective function would be:
\begin{equation}\label{eq:minimize}
\begin{split}
\textrm{$minimize_x$}~~\;cf(x,t) +\beta||y(x')-y(x_t)||_1+ ||x-x_0||_2 ^2\\
\textrm{such that }\;x  \in [0,1]^n\\
\end{split}
\end{equation}
where f(x,t) denotes the loss function and $\beta$ is regularization parameter for $L_1$ penalty. Increasing $\beta$ forces a lower $L_1$ distance between the adversarial and target benign and will result in same behavior as a legitimate image under noise for the generated adversarial.

\begin{figure}[htbp]
\centering
 \includegraphics[width=\linewidth]{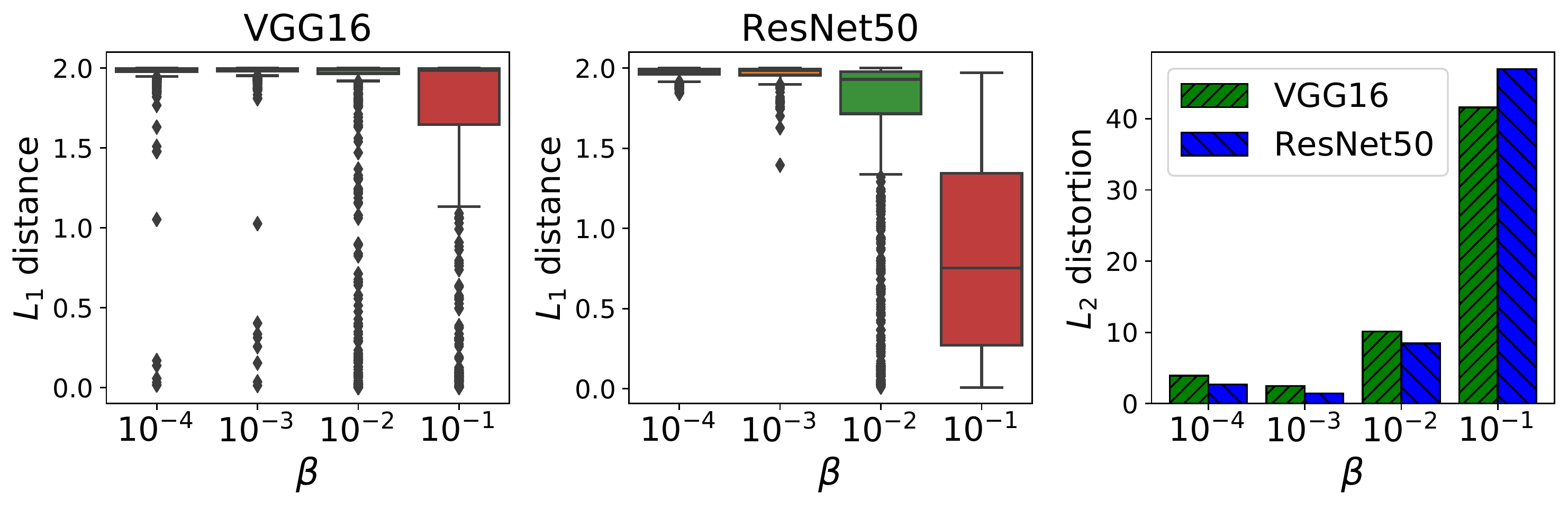}
\caption{\small $L_1$ distance vs. $L_2$ distortion for different $\beta$ values.}\label{Fig.:beta}
\end{figure}

Figure \ref{Fig.:beta} shows the effect of $\beta$ on the $L_1$ distance of probability distribution and $L_2$ distortion. Optimizing for both low $L_2$ distortion and $L_1$ distance are competing objectives. Increasing $\beta$ will minimize the $L_1$ distance but incur higher $L_2$ distortion. While low values of $\beta$ decrease the $L_1$ distance, the attack cannot lower it below HASI's detection threshold. On the other hand, increasing $\beta$ causes the $L_2$ distortion to surge also reduces the {\em attack} success rate. 

\begin{figure}[htbp]
\centering
 \includegraphics[width=\linewidth]{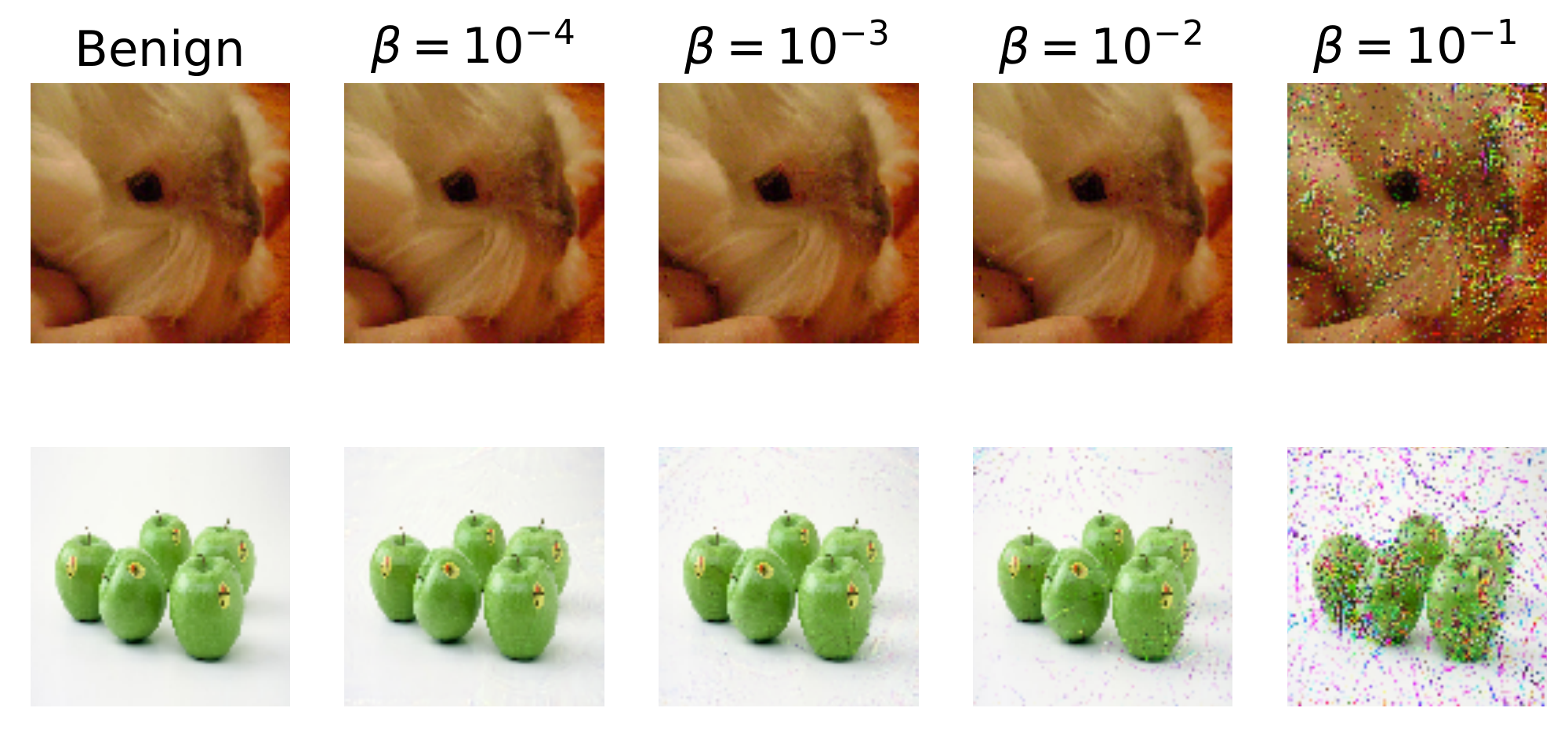}
\caption{\small Adversarials generated by HASI-aware attacks.}\label{Fig.:sample_adaptive}
\end{figure}

Figure \ref{Fig.:sample_adaptive} shows a couple of adversarial images generated with different $\beta$ values. Adversarials with $\beta=10^{-1}$ that can defeat HASI have noticeable perturbations and can be detected through other means. 

\subsection{Attack Strength Analysis}
\label{strength_analysis}
\begin{figure}[b]
  \centering
    \includegraphics[width=\linewidth]{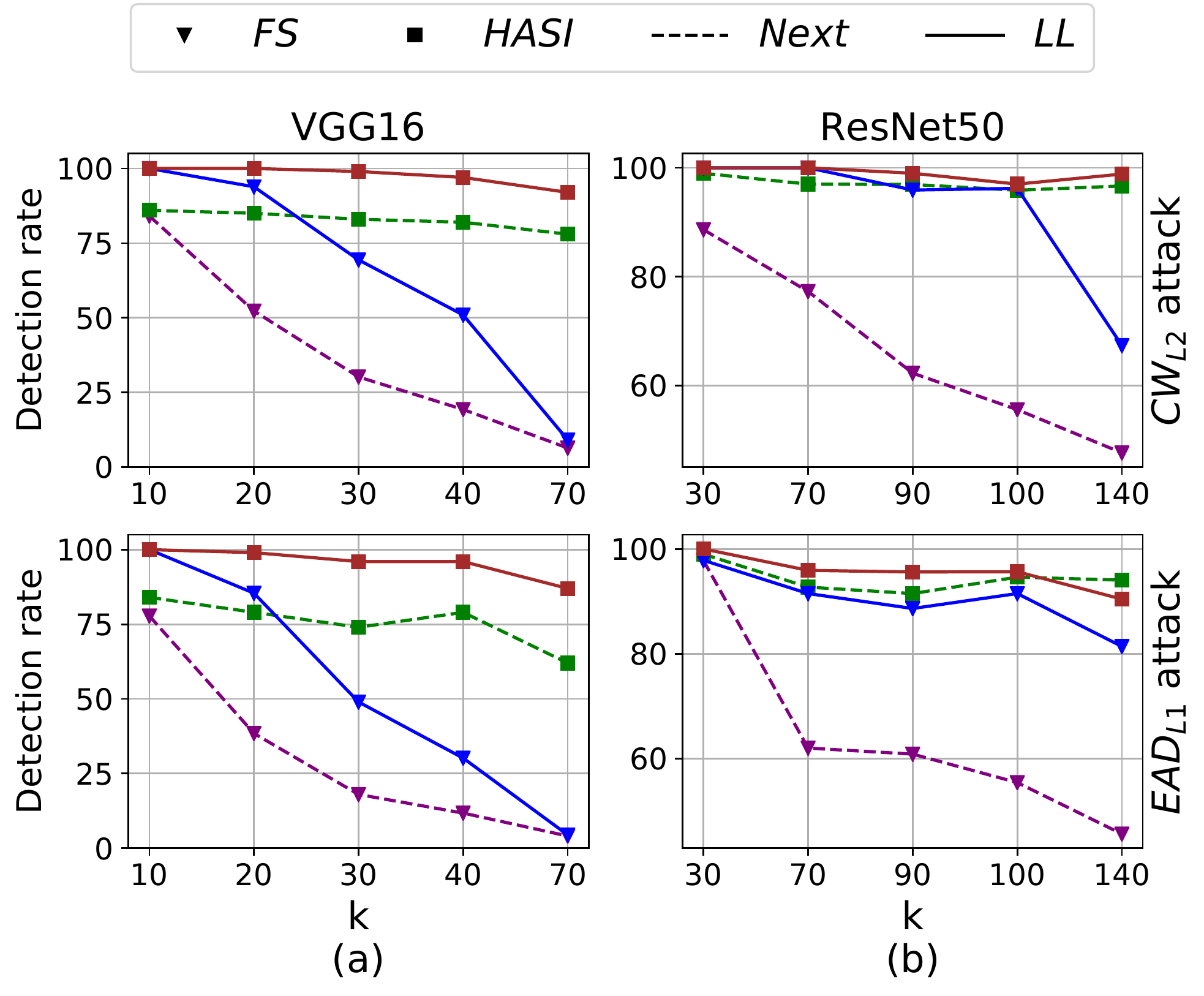}
  \caption{\small Detection rate for different k values and targets, (a) VGG16 and (b) ResNet50}\label{Fig.:strong_attack}
\end{figure}

Figure \ref{Fig.:strong_attack} further details the effect of adversarial strength on the adversarial detection rate for HASI and FS as a function of {\em k} for $CW$ and $EAD$ attack. 
Increasing the factor {\em k} (shown on the x-axes) improves the strength of the adversarials while maintaining the $L_2$ distortion in a reasonable range. For VGG16 the FS detection rate decreases to below 7\% and 5\% for CW and EAD while HASI maintains its detection rate above 78\% and 62\% for CW and and EAD respectively. For ResNet50, FS detection rate also degrades for higher k for EAD but HASI maintains detection rate above 90\%. 
The main reason for why HASI scales better with the strength of the attack is that HASI adapts the level of noise injection to the confidence of the classification. 

\end{document}